\documentclass[useAMS,usenatbib]{mnras}
\usepackage{amsmath}
\usepackage{graphicx}
\usepackage{xcolor}
\usepackage{hhline}
\usepackage{cancel}
\usepackage[normalem]{ulem}
\usepackage{amssymb}
\usepackage{amsfonts}
\usepackage{enumerate}
\usepackage{afterpage}
\usepackage{color}
\usepackage{multirow}
\usepackage[T1]{fontenc}                
\usepackage{ae,aecompl}
\usepackage{natbib}
\newcommand{\be}{\begin{equation}}
\newcommand{\ee}{\end{equation}}

\definecolor{darkblue}{RGB}{14,0,185}
\definecolor{darkred}{RGB}{175,0,0}

\newcommand{\class}{{\sc {class}}}

\begin{document}

\title[Cosmological Constraints with Clustering Redshifts]{Cosmological Constraints with Clustering-Based Redshifts}
\author[Ely D. Kovetz, Alvise Raccanelli and Mubdi Rahman]{Ely D. Kovetz, Alvise Raccanelli and Mubdi Rahman\\
Department of Physics and Astronomy, Johns Hopkins University, Baltimore, MD 21218 USA}

\maketitle
\label{firstpage}

\begin{abstract}
We demonstrate that observations lacking reliable redshift information, such as photometric and radio continuum surveys, can produce robust measurements of cosmological parameters when empowered by clustering-based redshift estimation. This method infers the redshift distribution based on the spatial clustering of sources, using cross-correlation with a reference dataset with known redshifts. 

Applying this method to the existing SDSS photometric galaxies, and projecting to future radio continuum surveys, we show that sources can be efficiently divided into several redshift bins, increasing their ability to constrain cosmological parameters. 
We forecast constraints on the dark-energy equation-of-state and on local non-gaussianity parameters. We explore several pertinent issues, including the tradeoff between including more sources versus minimizing the overlap between bins, the shot-noise limitations on binning, and the predicted performance of the method at high redshifts.  

Remarkably, we find that, once this technique is implemented, constraints on dynamical dark energy from the SDSS imaging catalog can be competitive with, or better than, those from the spectroscopic BOSS survey and even future planned experiments. Further, constraints on primordial non-Gaussianity from  future large-sky radio-continuum surveys can outperform those from the Planck CMB experiment, and rival those from future spectroscopic galaxy surveys. The application of this method thus holds tremendous promise for cosmology.
\end{abstract}


\section{Introduction}
The parameters of the concordance cosmological model have now been 
measured with increasing precision by three generations of cosmic microwave background (CMB) satellites. 
Meanwhile, the tremendous potential of the variety of multi-wavelength galaxy surveys remains hitherto largely untapped. 
To compete with the CMB, galaxy surveys need to reach high-redshifts over larger parts of the sky, where obtaining 
precise redshift information becomes increasingly difficult. While photometric and radio continuum 
surveys can access this regime more easily, this comes with the price of poor or absent source-redshift information.

In this work we study the potential improvement from using the clustering properties of galaxies in order to augment the photometric determination of their redshifts,
a concept first developed by \citet{1979ApJ...227...30S}, and expanded upon by various groups \citep{2013arXiv1303.4722M,2008ApJ...684...88N, 2013MNRAS.433.2857M}.
The fundamental observable of the clustering-based redshift (CBR) method is the angular cross-correlation of an \emph{unknown} sample, where redshift information is desired, with a series of consecutive slices of a \emph{reference} sample with known redshifts. The cross-correlation amplitude is related to the redshift distribution of the unknown sample and its bias factor, as discussed in \citet{2013arXiv1303.4722M} and \citet{2015MNRAS.447.3500R}. Using the prescribed approach, the degeneracy with the bias factor can be minimized for samples with narrow redshift distributions, and the overall effect on the inferred distribution is minimal for most practical samples. Consequently, one  infers the redshift distribution of a sample of galaxies, based only on the angular position of the sources. The availability of spectroscopic reference samples composed of galaxies and quasars up to $z \sim 6$, particularly in the Northern Hemisphere with the Sloan Digital Sky Survey (SDSS) spectroscopy, has enabled this technique to be used to infer the redshift distribution of the photometric sources in the SDSS \citep{2015arXiv151203057R}, the 2MASS near-infrared extended and point sources \citep{2016MNRAS.tmp...41R}, and even the cosmic infrared background from Planck \citep{2015MNRAS.446.2696S}. 

To study the benefits of this method in cosmology, we forecast constraints on parameters which are directly related to fundamental ingredients of the standard cosmological model, namely the nature of dark energy and the properties of inflation. The pressing question concerning dark energy is whether it is a simple cosmological constant or if it originates from a dynamical degree of freedom and thus varies in time. The latter scenario may lead to observable effects, especially when the Universe is dark-energy dominated, and thus galaxy surveys such as SDSS, reaching out to $z\sim1$, are potentially a much stronger probe than the CMB, provided that the redshift information is attainable. Meanwhile, an important test of inflation is the measurement of the deviation from Gaussianity of the primordial density fluctuations. While standard single-field slow-roll inflation predicts very small non-Gaussanities~\citep{Maldacena:2003}, multi-field models of inflation generically predict deviations that may be smaller than the strongest bound achievable by the CMB~\citep{Planck:NG}, but will be within reach of upcoming galaxy surveys once they outperform the CMB in terms of the number of measured modes. This can be accomplished by using deep radio surveys to probe higher redshifts, where more volume is accessible and more modes are in the linear regime, but in turn the challenge of acquiring reliable redshift information becomes daunting.

As our results demonstrate, empowered by CBR estimation, the constraints on these two models can vastly improve. When this method is implemented for the SDSS imaging catalog, we find that the forecasts for dark-energy constraints exceed those from spectroscopic surveys. We also consider large-sky radio-continuum surveys, where color information is scarce and simple photometry is very limited. Based on our predicted performance of the technique, we show that the resulting constraints on non-Gaussianity are highly competitive with those from alternative surveys. We conclude our analysis with a qualitative discussion of the limitation of binning.

\section{Data}
\label{sec:Data}

\subsection{Sloan Digital Sky Survey Photometric Galaxies}

The SDSS photometric survey remains the largest wide-field galaxy survey, covering one-third of the sky with over 200 million sources down to $r < 22$ with photometry in the $ugriz$-bands from 0.3 to 1 micron~\citep{2014ApJS..211...17A, 2000AJ....120.1579Y}. For low-redshift sources, these bands will be dominated by flux produced from stellar photospheres, providing significant variations in the spectral energy distributions of the sources across this wavelength region based on galaxy type and redshift. Consequently, it is possible to narrowly bin sources in redshift from this catalogue using photometry alone. We adopt the clustering redshift results from~\citet{2015arXiv151203057R}, where redshift distributions are inferred for sources, binned in each of the four possible colours, as well as all four colours together (using photometric redshift as a proxy). We also adopt the redshift-dependant bias of these galaxies as inferred from this paper. For this work, we choose samples of sources selected using photometric redshifts, and adopt the inferred redshift distribution for each sample using clustering redshifts.

\subsection{Radio Continuum Surveys}
Radio continuum surveys measure the integrated emission of radio sources in one broad frequency band. At radio wavelengths, the spectral energy distributions of most sources are generally smooth and featureless.
For this reason, there is little distance information to be gained by integrating the flux density over more than a single, wide band. In turn, using a wide band significantly increases sensitivity, allowing much fainter sources to be detected than would be possible otherwise, hence allowing continuum surveys to cover extremely large volumes, but at the expense of radial information.
In this work we consider a future experiment modeled after the proposed design of the Square Kilometre Array continuum survey (SKA,~\citealt{SKA:Jarvis}).
Given the uncertainties in the exact details of the actual SKA instrument that will be built, we forecast predictions by taking several flux limits from the $S^3$ simulation\footnote{\url{http://s-cubed.physics.ox.ac.uk}}, over an area of $3\pi$ steradians.
To determine the bias of these radio sources, we use the prescription described in~\cite{Wilman:2008}: each galaxy population is assigned a specific dark matter halo mass which is chosen to reflect the large-scale clustering found by observations; the bias is then simply the dark matter halo bias for that mass, computed following the formalism in~\citet{Mo:1996} (for more details on the redshift distribution and the bias of different populations, see also~\citealp{Raccanelli:radio}). We assume the bias to be scale-independent as we focus on large-scale correlations.

It is worth noting that while large uncertainties in the modeling of the bias remain, \citealt{Lindsay:2014b} show that at low redshifts the bias used for $S^3$ is in good agreement with recent observations.
For this reason, in computing our results we assume the redshift dependence of the bias is known. In the optimistic case, we assume the bias is known perfectly. To be more conservative, we also allow for a constant shift in the bias and consider two additional cases: marginalizing over it, and adopting a prior. One way to produce such a prior is through the cross-correlation of radio sources with CMB-lensing maps (e.g.~\citealt{Sherwin:2012, Das:2013}).

\section{Constraining Cosmological Parameters}

To constrain cosmological parameters, we forecast measurements of the two-point correlation function of source positions, which is
a measure of the degree of clustering distribution of sources. 
The angular power spectrum of galaxy correlations can be calculated from the underlying 3D matter power spectrum using
\begin{equation}
\label{eq:Clgg}
C_{\ell}^{gg} = 4 \pi\int \frac{dk}{k} \Delta^2(k) [W_{\ell}^g(k)]^2 \, 
\end{equation}
where $\Delta^2(k)$ is the matter power spectrum today, and $W_{\ell}^g$ is the source distribution window function, which can be written as (see e.g.~\citealt{Raccanelli:2008})
\begin{equation}
\label{eq:flg}
W_{\ell}^g(k) = \int \frac{dN}{dz} b(z) D(z) j_{\ell}[k\chi(z)] dz,
\end{equation}
where  $(dN/dz)dz$ is the mean number of sources per steradian with redshift
$z$ within $dz$, $b(z)$ is the bias factor relating the source overdensity to the mass overdensity, assumed to be scale-independent, $D(z)$ is the linear growth factor of mass fluctuations, $j_{\ell}(x)$ is the spherical
Bessel function of order $\ell$, and $\chi(z)$ is the comoving distance. For the sake of sanity in the notation, we drop the superscript ``gg" in the power spectrum in Eq.~(\ref{eq:Clgg}) in what follows.

The Fisher matrix, defined as the derivative with respect to pairs of model parameters of the log-likelihood, is given by
\be
\label{eq:fisher}
F_{\alpha \beta} = \sum \frac{\partial C_\ell^{(ij)} }{\partial \lambda_\alpha} \frac{\partial C_\ell^{(pq)}}{\partial \lambda_\beta} \sigma^{-2}_{C_{\ell \, \rm[(ij), (pq)]}},
\ee
where $\lambda_\alpha,\lambda_\beta$ are the parameters we wish to constrain and the covariance is calculated from the errors $\sigma_{C_\ell}$ in the power spectra. The latter are computed as (see e.g.~\citealt{Raccanelli:2015GR})
\begin{equation}
\label{eq:err-clgt}
\sigma^2_{C_{\ell \, \rm[(ij), (pq)]}} = \frac{\tilde{C}_{\ell}^{\rm (ip)} \tilde{C}_{\ell}^{\rm (jq)} + \tilde{C}_{\ell}^{\rm (iq)} \tilde{C}_{\ell}^{\rm (jp)}}{(2\ell+1)f_{\rm sky}} \, , ~~\tilde{C}_{\ell}  = C_{\ell}^{ij} + \frac{\delta_{ij}}{dN(z_i)/d\Omega} \,
\end{equation}
where $\tilde{C}_{\ell} $ is the observed power spectrum, including the shot noise.

Here $dN(z_i)/d\Omega$ is the average number of sources per steradian within the bin $z_i$.
We sum over the matrix indices $(ij)$ with $i\leq j$ and $(pq)$ with
$p \leq q$ which run from 1 to the number of bins.
This formalism ensures that both auto and cross-bin correlations are used with the full covariance correctly taken into account, which is particularly important in the case of overlapping bins. 

In practice, we compute the derivatives in Eq.~(\ref{eq:fisher}) numerically, using outputs from the \class{}\footnote{{\url{http://class-code.net/}}} code. We set the minimum angular scale for our analysis by fixing the maximum multipole to $\ell=200$.
Finally, when forecasting constraints, we consider three different scenarios with respect to the galaxy bias: marginalizing over a constant shift in its value, adopting a prior for this shift, and maximizing the likelihood with respect to it (i.e.~assuming it is perfectly known). 

Dynamical dark energy can be distinguished from a cosmological constant by considering the time evolution of the equation of state, $w = \frac{p}{\varrho}$, where $p$ and $\varrho$ are the pressure and energy density of the dark energy fluid, and for a cosmological constant, $w=-1$.
We adopt the widely used  parameterization~\cite{Linder:2003, Linder:2005}
\begin{align}
w(a) = w_0 + w_a (1-a) \, .
\end{align}
We thus expect to measure deviations from $w_0=-1$ and $w_a=0$ if dark energy is not sourced by a cosmological constant.

Primordial non-Gaussianity can be identified in large-scale structure observations by measuring the induced scale-dependent halo bias~\citep{Dalal:2008}
\begin{equation}
\label{eq:ng-bias}
\Delta b(z, k) = [b_{\rm G}(z)-1] f_{\rm NL}(k)\delta_{\rm c} \frac{3 \Omega_{m}H_0^2}{c^2k^2T(k)D(z)}, 
\end{equation}
where $f_{\rm NL}$ is the ``local type" non-Gaussianity, $b_{\rm G}(z)$ is the usual bias calculated assuming scale-independent Gaussian initial conditions, $D(z)$ is the linear growth factor, $\Omega_{m}$ and $H_0$ are the matter energy-density and Hubble parameter today, $\delta_{\rm c}\sim 1.68$ is the critical overdensity value for spherical collapse, and $T(k)$ is the matter transfer function. We shall examine how $f_{\rm NL}$ can be constrained using redshift-binned galaxy correlation measurements on large scales. 
We also consider the running of $f_{\rm NL}$, which describes its scale-dependent behavior that arises in some inflationary models ~\citep{Liguori:2006, Khoury:2009}. 
We parameterize this as~\citep{Shandera:2011, Raccanelli:fNL}
\begin{equation}
\label{eq:biasfNL}
f_{\rm NL}(k) = f^{\rm pivot}_{\rm NL} \left(\frac{k}{k_{*,\rm{NG}}}\right)^{n_{\rm NG}} \, ,
\end{equation}
where $k_{*,\rm{NG}}$ is the pivot scale, which we set to $0.04$ Mpc to facilitate comparison with CMB analyses. We set a fiducial $f^{\rm pivot}_{\rm NL}=30$.

\section{Results}
\label{sec:Results}

\subsection{Clustering-based Redshift Bins for SDSS and Radio Surveys}

A challenge of the CBR method is that it does not \emph{intrinsically} provide a redshift for each of the sources in the unknown sample, but rather a redshift distribution for the entire sample, or equivalently, a redshift probability distribution for any source within the sample. Consequently, to obtain narrower redshift distributions, we subdivide the sample using a variety of observable parameters of the source, including but not limited to flux and colour. We can produce clustering redshift distributions for each of these subsamples.

For SDSS, we use actual data from \citet{2015arXiv151203057R}, while for the future radio sources, we use simulated data from the S$^{3}$ extragalactic continuum simulation \citep{Wilman:2008}. 

\vspace{-0.1in}
\subsubsection{Single Bin Case}

In their most basic form, each catalogue contains a list of sources over a range of redshifts; while the angular positions are known, the redshift distribution in principle is not, and is expected to cover a wide range ($\Delta z \sim 0.8$ for SDSS, $\Delta z \sim 6$ for SKA-like sources). The simplest procedure to infer cosmological parameters from these sources is to use the full catalogue as a whole. To do this, however, one requires either an {\em a priori} assumption or inference of the catalogue's redshift distribution. Conveniently, the clustering redshift technique can provide this information, both in the case where the measured fluxes encode substantial redshift information (such as the SDSS galaxies), and in cases where they do not (such as radio continuum observations). 

However, with such wide redshift ranges, the inferred redshift distributions become more degenerate to the bias of the sample \citep[for details, see][]{2013arXiv1303.4722M}. This increases the uncertainty in the measured redshift distribution, but can be alleviated by subsampling the catalogue into slices with narrower redshift distributions. This is relatively simple for optical catalogues where galaxy colours strongly relate to redshift \citep{2015arXiv151203057R}, but becomes challenging in other wavelength regimes where redshift information is more difficult to extract on a per-object basis. In this case, it makes most sense to remove sources from the catalogue where all of the measured properties of the source (typically fluxes) are fully ambiguous with redshift, and using the remaining subset for the cosmological parameter inference. We discuss this in the next case. 

We emphasize that as a result of this process to identify a single redshift bin with a reliable redshift distribution, we often have to sacrifice some of the sources in the full catalogue. Therefore it is important to appreciate that in our analysis below, the comparison is between CBR-recovered single and multi bins.

\vspace{-0.1in}

\subsubsection{Multiple Bins Case}

\begin{figure}
\centering
\includegraphics[width=0.95\columnwidth]{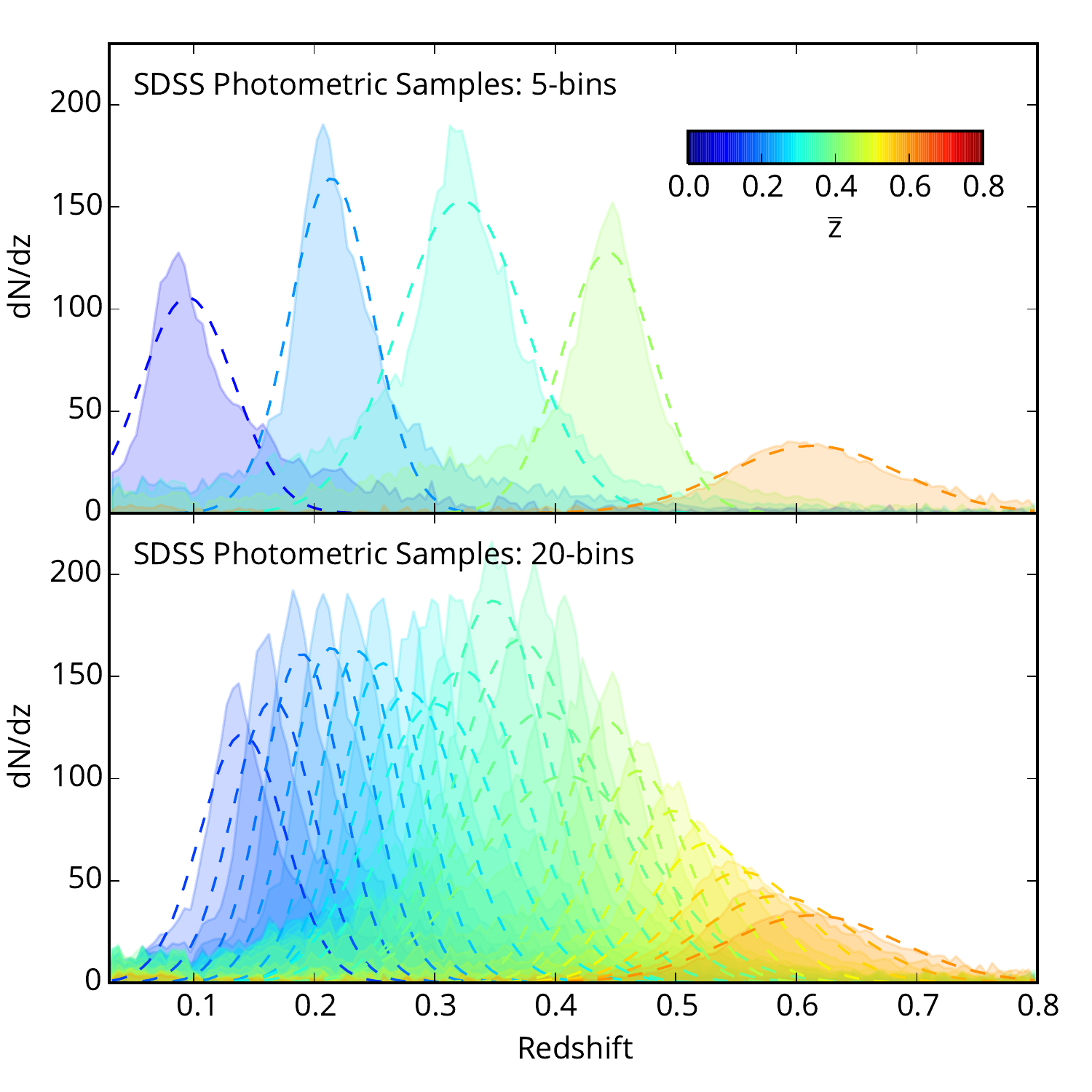}
\caption{Isolated samples selected using SDSS photometry, with redshift distributions inferred with clustering redshifts from Rahman et al. 2015. The shaded regions indicate the clustering redshift distribution of the samples, and the dashed lines show Gaussian fits to the distributions. The colour of the curves indicate the mean redshift of the distribution. The dN/dz is given as a total number of sources over the SDSS footprint. \emph{Top:} The 5-bin case with minimal overlap between samples. \emph{Bottom:} The 20-bin case, with many more sources included in total, but also substantial overlap between samples.}  
\label{fig:sdssbins}
\end{figure}

As mentioned earlier, we select SDSS bins (with $r < 22$) based on photometric redshifts, which are a proxy for the selection of sources in the four-dimensional colour space. \citet{2015arXiv151203057R} segment sources with photometric redshifts between 0.02 and 0.8 with 320 separate bins, for which clustering redshifts are available. However, the typical redshift width of the resultant bins is $\Delta z \sim 0.1$, indicating that adjacent bins have significant redshift overlap, rendering their use for cosmological constraints limited. Consequently, we select a series of five bins that minimize redshift overlap (Fig.~\ref{fig:sdssbins}, top panel), as well as a series of 20 bins with more significant overlap (Fig.~\ref{fig:sdssbins}, bottom panel).  Our Fisher analysis takes the effect of increasing overlap into account, as explained in the previous section. The typical number of sources in these bins range from $10^5$ to a few times $10^6$ sources. For the purposes of the analysis, we fit the derived clustering redshift distributions with gaussian profiles. 

For the simulated radio survey, we use simulations for an SKA-like survey with wavelength coverage from 151 MHz to 4.8 GHz. We choose sources down to the respective 1.4 GHz flux cut for each survey, a conservative 10$\mu$Jy for SKA at $5\sigma$ and a more optimistic 1$\mu$Jy. We assume inferred clustering redshift distributions, primarily from SDSS quasars, which have been used for clustering redshifts at similar redshifts \citep[e.g., ][]{2015MNRAS.446.2696S}. The challenge of radio observations is finding parameters on which to bin for narrow redshift distributions; radio emission, with smaller spectral variations as a function of source and/or redshift, will have substantially greater degeneracy in the flux ratios or ``radio colours'' of the sources. Consequently, a wider wavelength coverage of the survey provides additional discriminating power to isolate bins with narrow redshift distributions.

\begin{figure}
\includegraphics[width=0.95\columnwidth]{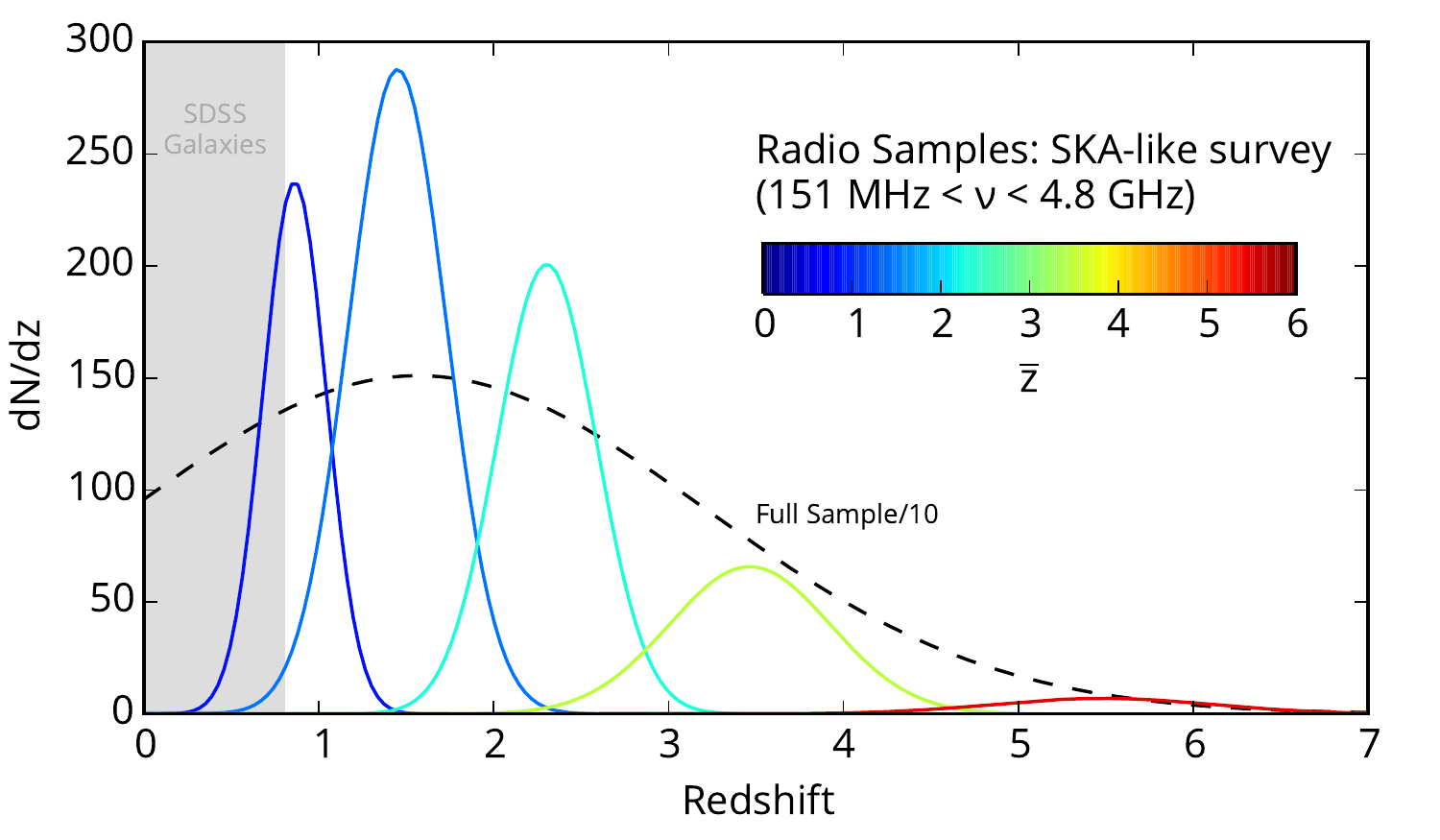}
\caption{Hypothetical redshift distributions for an SKA-like radio survey, as identified from the S-cubed simulation. These samples are separable using the frequency information from the chosen survey. The dN/dz is given as the number per square degree. The colour of the curve indicates the mean redshift of the distribution. The black dashed curve indicates the total galaxy distribution of all sources, divided by a factor of 10 to fit on the plot. The gray region indicates the redshifts probed by the SDSS galaxies from Fig.~\ref{fig:sdssbins}.}  
\label{fig:radiobins}
\end{figure}

To isolate samples, we segment the sources in three ``radio colours'', using the flux ratio of the 151 MHz, 700 MHz, 1.4 GHz and 4.8 MHz emissions. From this procedure, we identify five bins for the SKA-like survey with relatively low overlap (Figure \ref{fig:radiobins}). As before, we fit the redshift distributions to gaussians for the analysis. We note that additional information for redshift binning \emph{may} be available by cross-matching the sources to observations in other wavelength regimes; however this process places a selection bias on the sources used for the measurement. Further, we note that for these sources, spectrosopic and/or photometric redshift information will not be available; it is then important to stress that clustering redshifts provide the only model-independent redshift information. 

\subsection{Forecasted Constraints}

 In Table \ref{tab:Sloan} we provide the constraints on the dark energy parameters based on the SDSS imaging catalog for the different scenarios with respect to the galaxy bias. With a reasonable prior on the bias, our predicted constraints using the CBR method are roughly a factor of four better than the joint constraints that were generated from CMASS+CMB~\citep{Samushia:2013}, which rely on BOSS DR9 spectroscopic sources. When using 20 bins instead of 5, the improvement is limited to less than $10\%$, as a result of the large overlap between the bins. It is noteworthy that our results suggest an improvement upon forecasts for eBOSS~\cite{Zhao:eBOSS}.
The recent updated Euclid science forecasts~\cite{Amendola:euclidnew} suggest constraints on $\{w_0, w_a\}$ very similar to our 5 bins case assuming the bias is known, or the 20 bin case with a prior on the bias.

\begin{table*}
    \begin{tabular}{ccccc}
         \hline
         Case Study & Parameter &   Marginalized bias& Known bias & $0.1$ bias prior \\
            \hline \hline
     1 bin (110M total sources) & $w_0$& 1.15 & 0.65 & 0.67 \\
        located at $z=0.3\pm0.27$  & $w_a$&6.55  &  1.30 &1.66 \\
            \hline 
         \multirow{2}{*}{5 bins (21M total sources)}  & $w_0$& 0.241 &   0.066  & 0.071 \\
          & $w_a$&1.71  & 0.14  & 0.23 \\
            \hline       
         \multirow{2}{*}{20 bins (93M total sources)}  & $w_0$& 0.142 &   0.054  & 0.059 \\
          & $w_a$&0.96  & 0.11  & 0.21 \\
            \hline       
             \end{tabular}
    \caption{Forecast constraints on $w_0$ and $w_a$ for SDSS. Results are shown for 1, 5 and 20 CBR bins, for three cases: ignoring bias, marginalizing over the bias and adding a $0.1$ bias prior. The improvement from 1 to 5 bins is extreme, even though many sources were lost. Going to 20 bins, while incorporating many more sources into the analysis, yields diminshing returns due to bin overlap and as the shot noise in each bin is higher (see the discussion regarding Fig.~\ref{fig:LimitationsofBinning}).}
    \label{tab:Sloan}
\end{table*}

Meanwhile, in Table \ref{tab:SKA} we show the resulting constraints on the non-gaussianity parameters 
with the CBR-recovered redshift bins for our radio continuum survey, which includes $300{\rm M}$ sources 
in total with a $10\mu {\rm J}$ flux cut at $1.4\,{\rm GHz}$. Note that the single CBR bin includes only  
$\lesssim50\%$ of the total sources in the simulation, while the total number of sources in the 5 bins is 
less than $10\%$ of those in the single bin. Nevertheless, our results, even with one bin, significantly 
outperform those from Planck~\citep{Planck:NG} and forecasts for eBOSS and Euclid~\citep{Zhao:eBOSS, Amendola:euclidnew}. 
We also calculated the results when marginalizing over a different constant shift to the bias in each of the 5 bins. 
This results in a minor change for the $1\sigma$ constraint on $f_{\rm NL}$ ($1.15\longrightarrow1.17$) for the 
case of no running, and a slightly larger degradation of the constraints to $\{\sigma_{f_{\rm NL}},\sigma_{n_{\rm NG}}\}=\{15.80,0.22\}$ for the running case.

\begin{table*}
    \begin{tabular}{ccccc|ccc}
         \hline
       &  &\multicolumn{3}{c}{No Running ~($n_{\rm NG}\equiv0$)~~~~~}  & \multicolumn{3}{|c}{With Running  ~($n_{\rm NG}\neq0$)~~~~~} \\
\hline      Case Study  &Par. & Margin. bias& Known bias &  $0.01$  prior & Margin. bias& Known bias &  $0.01$ prior \\
            \hline \hline
1 bin, $S_{\rm 1.4\, GHz}<10\mu{\rm Jy}$ ($1\mu{\rm Jy}$) & $f_{\rm NL}$& 0.73 (0.59) & 0.72 (0.58) & 0.72 (0.58) & 22.26 (20.75) & 7.50 (7.68) &  9.34 (8.61)\\
   at $z=1.56\pm1.64$ ($1.99\pm1.82$)        & $n_{\rm NG}$&  - & - & - &0.31 (0.29) & 0.12 (0.12) &0.14 (0.13)\\
            \hline  
      \multirow{2}{*}{5 bins, $S_{\rm 1.4\, GHz}<10\mu{\rm Jy}$ ($1\mu{\rm Jy}$)}      & $f_{\rm NL}$& 1.15 (0.56) &  1.11 (0.55) & 1.13 (0.56) & 9.82 (7.36) & 8.06 (3.05) & 7.33 (5.87)\\
          & $n_{\rm NG}$ & -& - & - &0.14 (0.11) & 0.12 (0.052) & 0.11 (0.091)\\
            \hline
        \end{tabular}
    \caption{Forecasts for non-gaussianity constraints from our SKA-like survey, using CBR. Results are shown for $f_{\rm NL}$ with and without running ($n_{\rm NG}$), and for three cases: ignoring bias, marginalizing over the bias and adding a $0.01$ bias prior. We show results for flux cuts of $10\mu {\rm J}$ and $1\mu {\rm J}$ (in parentheses) at $1.4\,{\rm GHz}$. }
    \label{tab:SKA}
\end{table*}

Besides the issue of bin overlap, which is taken into account in our Fisher analysis, the obvious 
tradeoff when striving to increase the number of bins over the same redshift range is that it  
results in fewer remaining sources---both in each individual bin and overall--- which hinders the 
ability to overcome the shot noise contribution.

\begin{figure}
\centering
\includegraphics[width=\columnwidth]{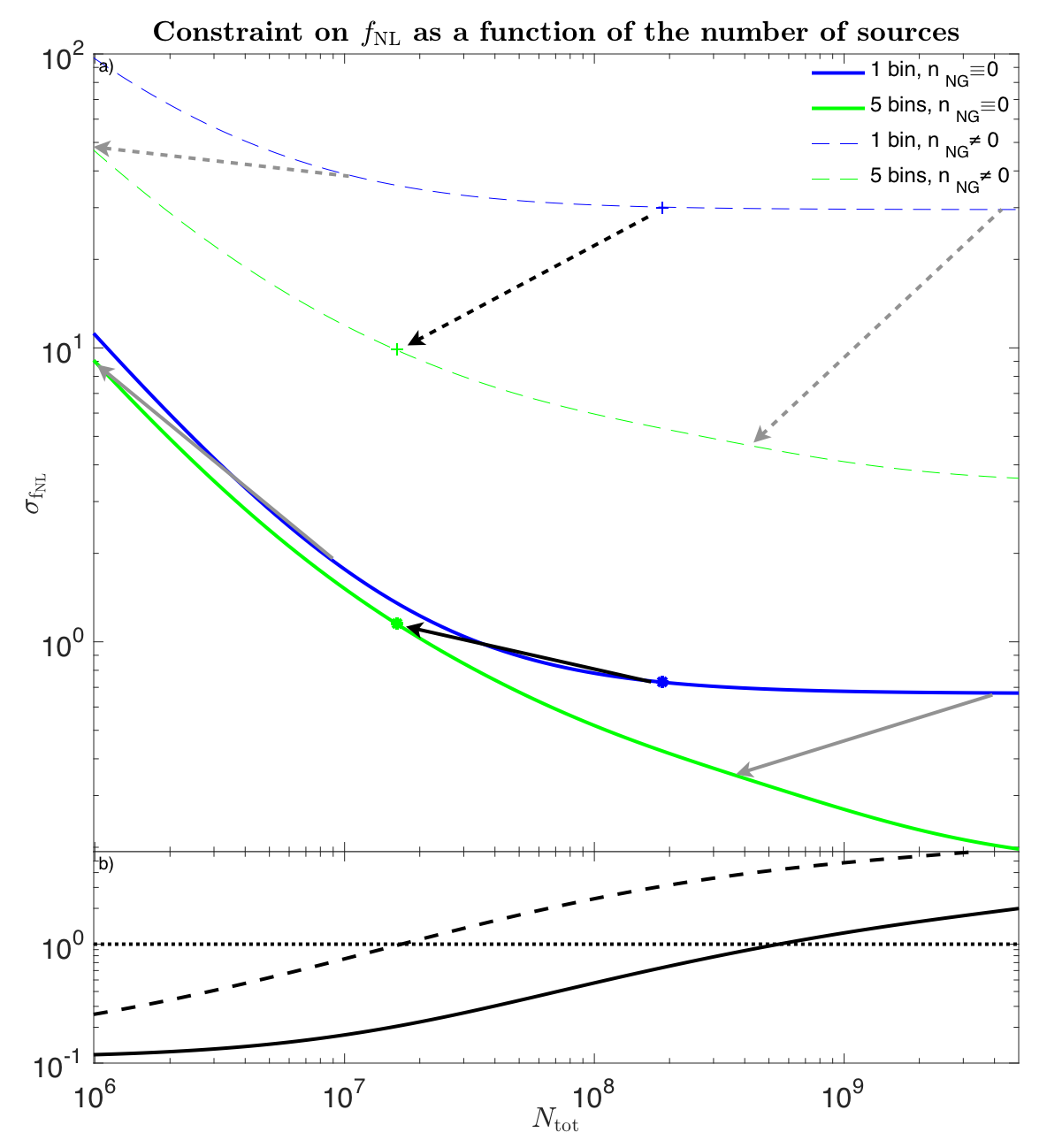}
\caption{The constraints on $f_{\rm NL}$ as a function of the total number of sources, for one and five bins. {\it Top:} Constraints in the no running scenario (solid lines) and allowing for running (dashed), both with marginalized bias. The $+$ and $\bullet$ symbols mark the constraints calculated specifically for our SKA-like survey (Table~\ref{tab:SKA}). {\it Bottom:} The ratio of the constraints for one and five bins, taking into account the loss of sources when using more bins. The arrows in the top panel indicate the effect---positive or negative--- of using more bins.}
\label{fig:LimitationsofBinning}
\end{figure}

To demonstrate this tradeoff, we show in the top panel of Fig.~\ref{fig:LimitationsofBinning} the 
$1\sigma$ constraint on $f_{\rm NL}$ as a function of the total number of radio sources in the bins, 
for the cases of one versus five CBR bins. The markers on the curves correspond to the results 
in Table \ref{tab:SKA} for our SKA-like survey in the marginalized bias scenario. The 5-bin 
constraint is based on $\lesssim10\%$ sources than the 1-bin case, and the curves are calculated 
under the assumption that this ratio holds throughout the range examined.
In the bottom panel we show, as a function of the number of sources in the single bin, 
the ratio between the constraints with 1 and 5 bins. It is evident, for example, that while the constraint 
on $f_{\rm NL}$ in the running scenario improves (see the dashed-black arrow which points downward), 
the improvement would be greater had the total number of sources in the one-bin case been larger, 
and that if it were significantly lower ($\lesssim10^7$), moving from 1 bin to 5 would result in a {\it worse} 
constraint, as indicated by the upward moving arrow in the left side of the plot. For the no-running case,
our survey is not yet in the regime where using more bins is beneficial.
While this analysis holds qualitatively for any experiment, when planning a specific observation it is important to 
examine this issue quantitatively given the conditions at hand.

\section{Conclusions}
\label{sec:Conclusions}

Looking forward, the role of the CMB as the primary source of constraining power for cosmological models is likely to be replaced by multi-wavelength galaxy surveys. In this work we explored the efficiency of the clustering-based redshift estimation technique, which uses information regarding the spatial clustering of sources to improve the photometric estimation of their redshifts, to enable redshift-binning for improved cosmological constraints.

The CBR technique is particularly valuable in light of the present landscape of large-scale surveys; spectroscopic redshifts, while precise, are particularly ``expensive'' to measure, and become more challenging to obtain with fainter sources. Further, their requirement of an identifiable feature in the spectra of the source, although common for low-redshift sources in the optical, becomes a problem in other wavelength regimes, such as radio and/or submilimeter where the spectral energy distributions of the sources tend to be smooth. 
Photometric redshifts, while being a valuable tool for optical observations at relatively low redshifts ($z < 1.5$), are subject to issues of ``catastrophic failures'', biases due to the model library and/or training set on which the algorithm is based, and become unreliable as the photometric accuracy decreases for fainter sources \citep{2010A&A...523A..31H}. Clustering redshifts, on the other hand, are robust to the problems of poor photometry and lack of spectral features since they only use information from the angular positions on the sky.
Additionally, the biases that arise in photometric redshifts from the training set of the algorithm have been shown to not affect the inferred redshift distribution from clustering \citep{2015MNRAS.447.3500R, 2015arXiv151203057R}. 

In this pioneering study we chose to focus on two distinctly different types of galaxy surveys---optical and radio continuum---and estimating the performance of CBR for these surveys, we addressed the measurement of the dark energy equation-of-state and primordial non-gaussianity parameters.  Applying CBR to the SDSS imaging catalog, our forecasts improve upon the constraints achieved using the CMASS BOSS catalog~\citep{Samushia:2013} and look to be competitive even with observations from future experiments such as eBOSS and the Euclid satellite~\citep{Zhao:eBOSS, Amendola:euclidnew}. Meanwhile, for an SKA-like radio survey, CBR will enable a measurement of non-gaussianity which handily beats the best constraint from the CMB, and rivals that of future surveys such as Euclid and SPHEREx~\citep{Amendola:euclidnew, SPHEREx}.

Naturally, there are lots of caveats to consider. We addressed here some of the most pressing ones, namely the cost of bin overlap and the reduction in the number of sources as the number of redshift bins is increased. The latter in particular, is an important tradeoff to take into account when considering a specific measurement. We showed how this may be done with a quantitative analysis for the case of non-gaussianity constraints with radio surveys, where we identified the minimum source number required for the binning to be useful. 
We stress that the constraints presented here are mere examples for illustrating the important role that CBR can play in cosmology. We chose to focus on the galaxy two-point correlation function, which has been extensively studied using other redshift-estimation techniques and therefore enabled a direct comparison of the resulting forecasts. A crucial issue for realistic measurements of cosmological parameters is the galaxy bias, which may be difficult to disentangle from variations in the cosmological parameters, although in general this is less worrisome in the linear regime, which the large-scale effects considered here are limited to. Indeed, our results when marginalizing over a constant shift in the bias value (over the full redshift range and in each bin separately) indicate a slight degradation of the constraining power if the bias is not well known. This issue may be overcome by using dedicated observables that are less sensitive to the bias (we plan to investigate this elsewhere).
Meanwhile, various combinations of probes or improvements in the modeling (such as including relativistic corrections, gravitational lensing, etc.), can result in better constraints than obtained here. 

What this work makes evident beyond doubt, however, is that there is tremendous promise for cosmology in acquiring redshift information based on the clustering properties of sources on the sky. 
Using CBR in cosmology, the sky is (indeed) the limit.

\vspace{0.5cm}
The authors thank Hiranya Peiris and especially Brice M\'enard for extended discussions. This work was supported by the John Templeton Foundation.

\vspace{-0.2in}

\bibliography{redshift_bins}

\bsp	
\label{lastpage}
\end{document}